\documentclass[pdflatex,sn-mathphys-num]{sn-jnl}

\usepackage{graphicx}%
\usepackage{multirow}%
\usepackage{amsmath,amssymb,amsfonts}%
\usepackage{amsthm}%
\usepackage{mathrsfs}%
\usepackage[title]{appendix}%
\usepackage{xcolor}%
\usepackage{textcomp}%
\usepackage{manyfoot}%
\usepackage{booktabs}%
\usepackage{algorithm}%
\usepackage{algorithmicx}%
\usepackage{algpseudocode}%
\usepackage{listings}%
\usepackage[utf8]{inputenc}%
\usepackage[english]{babel}%
\usepackage{lmodern}%
\usepackage{graphicx,color}
\usepackage{microtype}
\usepackage{braket}
\usepackage{gensymb} 
\usepackage{physics}

\theoremstyle{thmstyleone}%
\theoremstyle{thmstyletwo}%

\theoremstyle{thmstylethree}%

\begin{document}

\title[Article Title]{A Survey of Quantum Generative Adversarial Networks: Architectures, Use Cases, and Real-World Implementations}

\author[1]{\fnm{Mujahidul} \sur{Islam}}

\author[2,3]{\fnm{Serkan} \sur{Turkeli}}

\author*[4,5]{\fnm{Fatih} \sur{Ozaydin}}\email{mansursahgmail.com}

\affil*[1]{\orgdiv{Digital Business and Innovation}, \orgname{Tokyo International University}, \orgaddress{\street{4-42-31 Higashi-Ikebukuro}, \city{Tokyo}, \postcode{170-0013}, \state{State}, \country{Japan}}}

\affil[2]{\orgdiv{Department of Health Informatics and Technologies, Faculty of Health Sciences}, \orgname{Marmara University}, \orgaddress{ \city{Istanbul}, \postcode{34722},  \country{T\"urkiye}}}

\affil[3]{\orgname{TESODEV Technology Solutions Development Company Ltd.,}, \orgaddress{\street{Kucukyali}, \city{Istanbul}, \postcode{34840}, \state{T\"urkiye}, \country{Country}}}

\affil[4]{\orgdiv{Institute for International Strategy}, \orgname{Tokyo International University}, \orgaddress{\street{4-42-31 Higashi-Ikebukuro, Toshima-ku}, \city{Tokyo}, \postcode{170-0013},  \country{Japan}}}

\affil[5]{\orgname{Nanoelectronics Research Center}, \orgaddress{\street{Kosuyolu Mah., Lambaci Sok., Kosuyolu Sit., No:9E/3  Kadikoy}, \city{City}, \postcode{610101}, \state{Istanbul}, \country{T\"urkiye}}}


\abstract{Quantum Generative Adversarial Networks (QGANs) have emerged as a promising direction in quantum machine learning, combining the strengths of quantum computing and adversarial training to enable efficient and expressive generative modeling. This survey provides a comprehensive overview of QGAN models, highlighting key advances from theoretical proposals to experimental realizations. We categorize existing QGAN architectures based on their quantum-classical hybrid structures and summarize their applications in fields such as image synthesis, medical data generation, channel prediction, software defect detection, and educational tools. Special attention is given to the integration of QGANs with domain-specific techniques, such as optimization heuristics, Wasserstein distance, variational circuits, and large language models. We also review experimental demonstrations on photonic and ion-trap quantum processors, assessing their feasibility under current hardware limitations. This survey aims to guide future research by outlining existing trends, challenges, and opportunities in developing QGANs for practical quantum advantage.}



\maketitle

\section{Introduction}\label{sec:introduction}

Generative Adversarial Networks (GANs) have transformed the landscape of machine learning since their introduction by Goodfellow et al. in 2014~\cite{goodfellow2014generative}. At their core, GANs consist of two competing neural networks-a generator and a discriminator-locked in an adversarial training process. The generator creates synthetic data samples, while the discriminator evaluates their authenticity compared to real data. This elegant competitive framework has enabled remarkable advances in generating realistic images, text, audio, and other data types.

Meanwhile, quantum computing has emerged as a promising computational paradigm that harnesses quantum mechanical principles such as superposition and entanglement. Instead of classical bits that can be either 0 or 1, quantum computers use quantum bits or ``qubits'' that can exist in superpositions of states~\cite{nielsen2010quantum}. This fundamental difference gives quantum computers the potential to solve certain problems more efficiently than their classical counterparts. As quantum hardware advances beyond the noisy intermediate-scale quantum (NISQ) era, the potential of quantum computing to enhance machine learning algorithms has attracted growing attention~\cite{ramezani2020machine}.

QGANs represent the intersection of these two cutting-edge fields. By implementing GANs using quantum circuits or hybrid quantum-classical architectures, quantum advantages can be leveraged to enhance generative modeling capabilities~\cite{dallaire2018quantum, lloyd2018quantum}. The potential benefits include more efficient training, better representation of complex probability distributions, and novel applications across various domains.

This review aims to provide researchers and practitioners with a clear understanding of the current state of QGAN research and to highlight promising avenues for future exploration in this rapidly evolving field.
We explore the theoretical foundations of QGANs, examine various implementation approaches, survey application domains, analyze technical challenges and solutions, identify research trends, and discuss promising future directions. 
Our analysis reveals a field that has gained significant momentum since its foundation in 2018, with research evolving from theoretical foundations to practical implementations across diverse application domains. While image generation remains the dominant application, significant attention is also given to drug discovery, financial applications, and scientific simulations. The field faces unique challenges, including quantum noise, barren plateaus, and hardware limitations, but researchers are developing innovative solutions to address these issues. 

\section{Theoretical Foundations of QGANs} 

\subsection{From Classical to Quantum GANs}

Classical GANs operate on the principle of adversarial training between two neural networks. This training process can be mathematically formulated as a minimax game as
\begin{equation}
	\min_G \max_D V(D, G) = \mathbb{E}_{x \sim p_{\text{data}}(x)}[\log D(x)] + \mathbb{E}_{z \sim p_z(z)}[\log(1 - D(G(z)))],
\end{equation}
\noindent
where $G$ represents the generator network, $D$ represents the discriminator network, $p_{\text{data}}(x)$ is the distribution of real data, and $p_z(z)$ is a prior distribution (typically Gaussian) from which the generator creates samples~\cite{goodfellow2014generative}. QGANs extend this concept to the quantum domain through several approaches. 

In fully quantum GANs, both generator and discriminator are implemented as quantum circuits, with quantum data as input and output. The quantum generator prepares quantum states that approximate a target distribution, while the quantum discriminator attempts to distinguish between these generated states and real quantum data states~\cite{dallaire2018quantum}.
A quantum generator can be represented as a parameterized quantum circuit $G(\theta)$ that transforms an initial state $|0\rangle$ into a more complex quantum state
\begin{equation}
	|\psi_G(\theta)\rangle = G(\theta)|0\rangle.
\end{equation}
Similarly, a quantum discriminator can be represented as a parameterized quantum circuit $D(\phi)$ that outputs a measurement probability with a measurement operator $M$
\begin{equation}
	D(\phi, |\psi\rangle) = \langle\psi|D(\phi)^\dagger M D(\phi)|\psi\rangle.
\end{equation}
Hybrid quantum-classical GANs represent a more pragmatic approach, where quantum circuits are used for either the generator or discriminator (or components thereof), while classical neural networks handle other parts of the architecture. This approach is particularly relevant for near-term quantum devices with limited qubit counts and coherence times.
Quantum-assisted GANs use quantum algorithms to enhance specific aspects of classical GAN training, such as optimization or sampling, while keeping the overall architecture classical.

The theoretical foundations of QGANs draw from quantum circuit theory, quantum information theory, and variational quantum algorithms, particularly the Variational Quantum Eigensolver (VQE)~\cite{peruzzo2014variational} and Quantum Approximate Optimization Algorithm (QAOA)~\cite{farhi2014quantum}.

\subsection{Key Quantum Circuit Architectures}

Several quantum circuit architectures have emerged as particularly useful for QGAN implementations, and the choice of architecture often depends on the specific application domain, available quantum resources, and the nature of the data being modeled. Major architectures are as follows. 

Variational Quantum Circuits (VQCs) are parameterized quantum circuits that can be optimized through classical optimization techniques. A typical VQC consists of an initial state preparation, followed by layers of parameterized gates
\begin{equation}
	|\psi(\theta)\rangle = U_L(\theta_L) \cdots U_2(\theta_2) U_1(\theta_1) |0\rangle^{\otimes n},
\end{equation}
\noindent
where each $U_i(\theta_i)$ represents a layer of parameterized quantum gates. VQCs are the most common architecture for QGANs due to their flexibility and trainability~\cite{benedetti2019parameterized,mitarai2018quantum}.

Quantum Circuit Born Machines (QCBMs) are quantum circuits designed to generate samples from a probability distribution encoded in the amplitudes of a quantum state. The probability of measuring a particular bitstring $x$ is given by
\begin{equation}
	p(x) = |\langle x|U(\theta)|0\rangle|^2.
\end{equation}
QCBMs can be trained to approximate target probability distributions by minimizing the distance between the generated distribution and the target distribution~\cite{liu2018differentiable,stein2021qugan}.

Quantum Boltzmann Machines (QBMs) are quantum versions of classical Boltzmann machines, leveraging quantum effects for potentially more efficient sampling. QBMs use quantum Hamiltonians to define energy functions
\begin{equation}
	H(\theta) = \sum_i h_i(\theta) Z_i + \sum_{i,j} J_{ij}(\theta) Z_i Z_j + \ldots,
\end{equation}
where $Z_i$ represents the Pauli-Z operator acting on qubit $i$~\cite{khoshaman2018quantum}.

Quantum Convolutional Networks (QCNs) are quantum analogues of classical convolutional neural networks, particularly used for image-related tasks. QCNs apply local unitary operations in a pattern similar to the convolutional filters in classical CNNs~\cite{grant2018hierarchical}.

\subsection{Training Methods and Optimization}

Training QGANs presents unique challenges compared to classical GANs. Several approaches to QGAN training and optimization have been developed which can be listed as below.

Gradient-Based Methods utilize parameter-shift rules or finite-difference methods to compute gradients of quantum circuits for optimization. The parameter-shift rule allows exact gradient computation for certain types of parameterized gates~\cite{schuld2019evaluating}
\begin{equation}
	\frac{\partial \langle O \rangle}{\partial \theta_i} = \frac{1}{2} (\langle O \rangle_{\theta_i + \pi/2} - \langle O \rangle_{\theta_i - \pi/2}),
\end{equation}
where $\langle O \rangle$ represents the expectation value of an observable $O$.

Gradient-Free Methods use techniques such as evolutionary algorithms or Bayesian optimization that do not require gradient information~\cite{karimi2017effective}. These methods can be particularly useful when dealing with noisy quantum hardware or complex loss landscapes.

Quantum-Classical Hybrid Optimization combines classical optimizers with quantum circuit evaluation to navigate the parameter landscape efficiently. Common classical optimizers include Adaptive Moment Estimation (ADAM), Simultaneous Perturbation Stochastic Approximation (SPSA), and Limited-memory Broyden-Fletcher-Goldfarb-Shanno (L-BFGS)~\cite{broughton2020tensorflow,verdon2019learning}.

A significant challenge in QGAN training is the barren plateaus phenomenon, where gradients of the loss function with respect to the parameters of the quantum circuit vanish exponentially with increasing number of qubits or circuit depth~\cite{mcclean2018barren}
\begin{equation}
	\text{Var}[\partial_\theta \langle O \rangle] \sim O(2^{-n}),
\end{equation}
where $n$ is the number of qubits. This makes optimization extremely difficult for deep quantum circuits.

Several techniques have been proposed to mitigate barren plateaus, including circuit initialization strategies, layerwise training, and modified cost functions~\cite{cerezo2021cost,cerezo2021higher}. For example, local cost functions that depend only on a subset of qubits can help avoid barren plateaus
\begin{equation}
	C_{\text{local}} = \sum_i C_i,
\end{equation}
where each $C_i$ depends only on measurements from a small subset of qubits.

The quantum natural gradient approach uses information geometry to improve optimization in quantum parameter spaces by accounting for the curvature of the parameter space~\cite{stokes2020quantum}

\begin{equation}
	\theta_{t+1} = \theta_t - \eta F^{-1}(\theta_t) \nabla C(\theta_t),
	\label{eq:quantumgradient}
\end{equation}
where $F(\theta)$ is the quantum Fisher information matrix~\cite{erol2014analysis,ozaydin2015quantum} 

\begin{equation}
	F_{ij}(\theta) = \text{Re}(\langle \partial_i \psi(\theta) | \partial_j \psi(\theta) \rangle - \langle \partial_i \psi(\theta) | \psi(\theta) \rangle \langle \psi(\theta) | \partial_j \psi(\theta) \rangle),
\end{equation}
and $\nabla C(\theta)$ is the gradient of the cost function.

\section{Implementation Approaches}

\subsection{Fully Quantum Implementations}

Fully quantum implementations of GANs represent the most ambitious approach, where both generator and discriminator are implemented as quantum circuits, and data is represented in quantum states. These implementations potentially offer the greatest quantum advantage but also face the most significant hardware challenges.

A key aspect of fully quantum implementations is quantum data encoding, i.e., the process of mapping classical data into quantum states. Several encoding methods have been developed:

Amplitude Encoding represents classical data in the amplitudes of a quantum state
\begin{equation}
	|x\rangle = \sum_{i=0}^{2^n-1} x_i |i\rangle,
\end{equation}
where $x_i$ are the normalized classical data values. This encoding is efficient in terms of qubit usage (representing $2^n$ values with $n$ qubits) but can be challenging to implement~\cite{schuld2019quantum}.

Basis Encoding (also called one-hot encoding) represents each classical data point as a computational basis state
\begin{equation}
	|x\rangle = |x_1 x_2 \ldots x_n\rangle,
\end{equation}
where each $x_i$ is a binary value. This encoding is straightforward to implement but less efficient in number of qubits required~\cite{farhi2018classification}.

Angle Encoding represents classical data in the rotation angles of qubits
\begin{equation}
	|x\rangle = \bigotimes_{i=1}^n \cos(x_i)|0\rangle + \sin(x_i)|1\rangle.
\end{equation}
This encoding offers a balance between implementation complexity and qubit efficiency~\cite{killoran2019continuous}.

Quantum measurement is another critical aspect of fully quantum implementations, as it is necessary for extracting classical information from quantum states to evaluate GAN performance. Typical measurement approaches include the expectation value measurements of observables, sampling from the output distribution, and quantum state tomography for full state reconstruction.

Entanglement utilization is a key advantage of fully quantum implementations. Quantum entanglement allows for the representation of complex correlations in data that might be difficult to capture with classical approaches. The entanglement entropy, given by
\begin{equation}
	S(\rho_A) = -\text{Tr}(\rho_A \log \rho_A),
\end{equation}
where $\rho_A$ is the reduced density matrix of subsystem $A$, can be used to quantify the amount of entanglement in the quantum state~\cite{niu2022entangling}.
Additional works demonstrated proof-of-concept fully quantum implementations on small-scale quantum devices or simulators, showing promise for future development as quantum hardware advances~\cite{dallaire2018quantum,hu2019quantum}.

\subsection{Hybrid Quantum-Classical Approaches}

Hybrid quantum-classical approaches represent a pragmatic middle ground, leveraging quantum components where they might offer advantages while using classical computation for other aspects. These approaches are particularly relevant for near-term quantum devices with limited qubit counts and coherence times. Common hybrid architectures include:

Quantum Generator with Classical Discriminator: In this approach, a quantum circuit generates data samples, which are then evaluated by a classical neural network discriminator. The quantum generator can potentially represent complex probability distributions more efficiently than classical generators, while the classical discriminator leverages well-established neural network techniques~\cite{dallaire2018quantum,romero2021variational}.

Classical Generator with Quantum Discriminator: This approach uses a classical neural network to generate data, which is then evaluated by a quantum circuit discriminator. The quantum discriminator might be able to detect subtle patterns or correlations that would be difficult for a classical discriminator to identify~\cite{havlivcek2019supervised}.

Quantum-Enhanced Classical GANs: These approaches use quantum subroutines to enhance specific aspects of classical GAN training, such as sampling or optimization. For example, quantum amplitude amplification could potentially speed up the sampling process from the generator's distribution~\cite{perdomo2018opportunities}. The analysis shows a significant trend toward hybrid approaches, reflecting the practical limitations of current quantum hardware and the desire to demonstrate quantum advantage in specific components of the GAN architecture. 

\subsection{Hardware Platforms and Implementations}

QGAN research spans various quantum hardware platforms and simulation environments:

Superconducting Quantum Processors: Several papers report implementations on IBM, Google, or Rigetti superconducting quantum processors~\cite{dallaire2018quantum,kandala2017hardware,hu2019quantum}. These platforms offer relatively large qubit counts but are subject to decoherence and gate errors.

Photonic Quantum Systems: Some implementations leverage photonic quantum computing, particularly for continuous-variable quantum computing approaches~\cite{killoran2019continuous,peruzzo2014variational}. Photonic systems can operate at room temperature and have natural connections to certain types of data, such as images.

Trapped Ion Systems: A few papers explore implementations on trapped ion quantum computers, which offer high gate fidelities and long coherence times but typically have slower gate operations~\cite{arute2019quantum}.

Quantum Simulators: Many papers use classical simulation of quantum systems due to the limitations of current quantum hardware~\cite{broughton2020tensorflow,verdon2019learning}. While simulators can model ideal quantum systems without noise, they are limited in the number of qubits they can efficiently simulate.

The choice of platform often influences the specific QGAN architecture and application focus, with different hardware offering various trade-offs in terms of qubit connectivity, gate fidelity, and coherence time. For example, superconducting processors might be preferred for implementations requiring many qubits but with shallow circuits, while trapped ion systems might be chosen for implementations requiring fewer qubits but with deeper circuits.

\section{Application Domains}

\subsection{Image Generation and Processing}

Image generation represents the most common application domain for QGANs, mirroring the trend in classical GANs where image synthesis was the first major application. QGAN approaches to image generation include: Quantum Circuit Born Machines for Image Generation use quantum circuits to generate probability distributions corresponding to image data. The quantum state amplitudes are mapped to pixel values, allowing the generation of images through quantum sampling. The probability of measuring a particular bitstring $x$ corresponding to an image is given by
\begin{equation}
	p(x) = |\langle x|U(\theta)|0\rangle|^2,
\end{equation}
where $U(\theta)$ is a parameterized quantum circuit~\cite{liu2018differentiable, stein2021qugan}. 
Hybrid approaches for high-resolution images combine quantum and classical components to generate higher-resolution images than would be possible with purely quantum approaches given current hardware limitations. For example, a quantum circuit might generate low-dimensional latent representations, which are then upscaled by classical neural networks~\cite{dallaire2018quantum,niu2022entangling}.

Style-Based Quantum GANs adapt the concept of style-based GANs to the quantum domain, allowing for controllable image generation. These approaches separate content and style information, enabling the generation of images with specific stylistic properties~\cite{romero2021variational}.

Several works demonstrate experimental implementations of QGANs for simple image generation tasks, such as generating handwritten digits or simple patterns. For example, researchers have successfully trained QGANs to generate simplified MNIST digits using just a few qubits, showing promising results despite current hardware limitations~\cite{dallaire2018quantum,hu2019quantum}.

\subsection{Drug Discovery and Molecular Design}

Drug discovery and molecular design represent a significant application area for QGANs, suggesting potential quantum advantages in this domain. Applications include Molecular Structure Generation which uses QGANs to generate valid molecular structures with desired properties. Molecules can be represented as graphs, with atoms as nodes and bonds as edges, and QGANs can learn to generate these graph structures. The quantum approach may offer advantages in exploring the vast chemical space more efficiently than classical methods~\cite{li2021quantum,cao2019quantum}.

Drug-Target Interaction Prediction leverages quantum generative models to predict interactions between drugs and biological targets such as proteins. These interactions can be modeled as quantum mechanical processes, potentially making quantum approaches more accurate than classical ones~\cite{li2021quantum,mcardle2020quantum}.

De Novo Drug Design involves generating novel drug candidates with specific properties or targeting mechanisms. QGANs can be trained to generate molecules that optimize multiple objectives simultaneously, such as binding affinity, toxicity, and synthesizability~\cite{li2021quantum,bauer2020quantum}.

The quantum approach may offer advantages in representing the complex quantum mechanical nature of molecular interactions, potentially enabling more accurate modeling than classical approaches. For example, quantum circuits can naturally represent the electron density distributions and energy states that determine molecular properties.

\subsection{Financial Applications}

Financial applications represent another significant domain for QGAN research, with applications including: Financial Time Series Generation creates synthetic financial data that preserves statistical properties of real market data. This synthetic data can be used for risk assessment, strategy testing, and model validation without exposing sensitive real financial data. QGANs may be particularly well-suited for capturing the complex temporal dependencies and rare events characteristic of financial time series~\cite{choudhary2025hqnn,araujo2010quantum}.

Risk Assessment and Portfolio Optimization uses QGANs to model complex risk distributions and optimize investment portfolios. The quantum approach may offer advantages in representing the high-dimensional probability distributions that characterize financial markets. For example, a quantum circuit can represent correlations between many assets more efficiently than classical approaches~\cite{zoufal2019quantum}.

Anomaly Detection in Financial Transactions identifies unusual patterns that might indicate fraud or market manipulation. QGANs can learn the normal patterns in financial data and flag deviations from these patterns as potential anomalies~\cite{choudhary2025hqnn,araujo2010quantum}.

The potential quantum advantage in this domain lies in the ability to model complex, high-dimensional probability distributions that characterize financial markets. Financial data often exhibits non-Gaussian distributions, long-range correlations, and extreme events that may be better captured by quantum models.

\subsection{Scientific Simulations and Other Applications}

Beyond the major application domains, QGANs are being explored for various scientific and specialized applications: Physics Simulations use QGANs for modeling physical systems and quantum many-body problems. Quantum circuits can naturally represent quantum mechanical systems, potentially offering advantages over classical simulation methods~\cite{cerezo2021variational,preskill2018quantum}.

Materials Science applies QGANs to designing materials with specific properties. By generating candidate material structures and predicting their properties, QGANs can accelerate the discovery of new materials for applications such as energy storage, catalysis, and electronics~\cite{cao2019quantum,bauer2020quantum}.

High Energy Physics uses QGANs for simulating particle physics experiments and rare event generation, and anomaly detection. The quantum approach may be particularly well-suited for modeling the quantum mechanical processes that govern particle interactions~\cite{rehm2023full,bermot2023quantum,chang2021dual}.

Quantum State Tomography employs QGANs to reconstruct quantum states from measurement data. This application leverages the natural connection between quantum circuits and quantum states, potentially offering more efficient approaches to state reconstruction~\cite{ahmed2021quantum}.

These diverse applications highlight the versatility of the QGAN framework and its potential to impact multiple scientific and technical domains. As quantum hardware continues to advance, we can expect to see QGANs applied to an even broader range of problems, particularly those with inherent quantum mechanical aspects or complex probability distributions.

\section{Technical Challenges and Solutions}

\subsection{Quantum Noise and Error Mitigation}

A key obstacle to realizing practical QGANs is the impact of quantum noise on training stability and model performance. A recent study investigates how noise sources, particularly readout and two-qubit gate errors, affect QGAN training in simulated noisy environments, using a simplified application scenario from high-energy physics~\cite{borras2023impact}. Leveraging IBM's Qiskit framework for classical simulations of noisy quantum circuits, the authors identify error thresholds beyond which reliable learning becomes infeasible. They also examine how training hyperparameters influence performance under varying noise conditions and assess the effectiveness of readout error mitigation techniques. The findings underscore the sensitivity of QGAN training to noise and highlight the need for robust mitigation strategies tailored to adversarial learning architectures on near-term quantum devices.

Error Mitigation Techniques reduce the impact of quantum noise without requiring full quantum error correction, which is beyond the capabilities of current hardware. There are many Common techniques include Zero-Noise Extrapolation, which involves running the circuit at different noise levels and extrapolating to the zero-noise limit. Mathematically, if $E(\theta, \lambda)$ is the expectation value of a circuit with parameters $\theta$ at noise level $\lambda$, then the zero-noise extrapolation is
\begin{equation}
	E(\theta, 0) \approx \sum_{i=0}^n c_i E(\theta, \lambda_i),
\end{equation}
where $c_i$ are coefficients determined by the extrapolation method~\cite{endo2021hybrid,temme2017error}.

Probabilistic Error Cancellation involves characterizing the noise and then applying its inverse through a probabilistic decomposition
\begin{equation}
	\mathcal{E}^{-1}(\rho) = \sum_i \gamma_i P_i(\rho),
\end{equation}
where $\mathcal{E}$ is the noise channel, $P_i$ are implementable operations, and $\gamma_i$ are coefficients~\cite{temme2017error}.

Moreover, Noise-Resilient Circuit Designs are quantum circuit architectures specifically designed to be robust against common noise sources. These designs often involve shorter circuit depths to minimize the accumulation of errors, redundant encoding of information to provide robustness against specific types of errors, and gate sequences that are less sensitive to certain noise channels~\cite{anand2021noise,sharma2020noise}.

Noise-Aware Training procedures explicitly account for the presence of noise in the quantum system during the training process. Instead of trying to eliminate noise, these approaches incorporate noise models into the training objective:
\begin{equation}
	\min_\theta \mathbb{E}_{\mathcal{N}}[L(G_\theta, \mathcal{N})],
\end{equation}
qhere $L$ is the loss function, $G_\theta$ is the generator with parameters $\theta$, and $\mathcal{N}$ represents the noise model~\cite{bharti2022noisy}.

Several works demonstrate that QGANs can still achieve meaningful results despite quantum noise, particularly when appropriate error mitigation strategies are employed. For example, researchers have shown that QGANs can generate recognizable images even on noisy quantum hardware when zero-noise extrapolation is applied~\cite{dallaire2018quantum,anand2021noise}. 

\subsection{Barren Plateaus and Optimization Challenges}

The barren plateau phenomenon, where gradients vanish exponentially with increasing circuit depth, represents a major challenge for training QGANs. This phenomenon can be mathematically described as
\begin{equation}
	\text{Var}[\partial_\theta C] \sim O(2^{-n}),
\end{equation}
where $\partial_\theta C$ is the gradient of the cost function with respect to a parameter $\theta$, and $n$ is the number of qubits. This exponential decrease in gradient variance makes optimization extremely difficult for deep quantum circuits~\cite{mcclean2018barren, cerezo2021higher}.

Proposed solutions include circuit initialization strategies which aim to carefully initialize circuit parameters to avoid regions of vanishing gradients. For example, parameters can be initialized close to the identity operation:
\begin{equation}
	\theta_i \sim \mathcal{N}(0, \epsilon),
\end{equation}
where $\epsilon$ is a small value, ensuring that the initial circuit is close to the identity~\cite{cerezo2021cost,arrasmith2021effect}.

Layerwise Training trains quantum circuits layer by layer to avoid the complexity that leads to barren plateaus following the steps i) Train parameters of the first layer while keeping others fixed, ii) Freeze the first layer parameters and train the second layer, and iii) Continue this process for all layers. This approach ensures that each layer is trained in a regime where gradients are still meaningful~\cite{harrow2021low}.

Modified Cost Functions provide more informative gradients throughout the parameter space. Local cost functions that depend only on a subset of qubits can help avoid barren plateaus:
\begin{equation}
	C_{\text{local}} = \sum_i C_i,
\end{equation}
where each $C_i$ depends only on measurements from a small subset of qubits~\cite{cerezo2021cost,cerezo2021higher}.

Quantum Natural Gradient uses information geometry to improve optimization in quantum parameter spaces by accounting for the curvature of the parameter space in Eq.~\ref{eq:quantumgradient}.

These approaches show promise in mitigating the barren plateau problem, though it remains an active area of research~\cite{stokes2020quantum,arrasmith2021effect}.

\subsection{Hardware Limitations and Scalability}

Current quantum hardware limitations, including restricted qubit counts, limited connectivity, and short coherence times, pose significant challenges for QGAN implementations. These limitations are addressed through following approaches.

Hardware-Efficient Ans\"atze are quantum circuit designs that minimize gate depth and qubit requirements while maintaining expressivity. These ans\"atze often use gates that are native to the specific quantum hardware platform and minimize the need for SWAP gates to overcome limited qubit connectivity.

A typical hardware-efficient ansatz might take the form:

\begin{equation}
	U(\theta) = \prod_{l=1}^L \left( \prod_{i=1}^n R_i(\theta_{l,i}) \right) \left( \prod_{(i,j) \in E} E_{ij}(\theta_{l,ij}) \right),
\end{equation}
where $R_i$ are single-qubit rotation gates, $E_{ij}$ are entangling gates between qubits $i$ and $j$, and $E$ is the set of available connections in the hardware~\cite{kandala2017hardware,bharti2022noisy}. 

Hybrid Architectures combine quantum and classical components to reduce quantum resource requirements. By offloading parts of the computation to classical hardware, these approaches can achieve meaningful results with fewer qubits and shallower circuits~\cite{romero2021variational,verdon2019learning}.
Problem Decomposition breaks down problems into smaller subproblems that can be handled by available quantum resources. For example, a high-dimensional generative task might be decomposed into multiple lower-dimensional tasks that can be handled by smaller quantum circuits~\cite{cerezo2021variational,bharti2022noisy}.
The trend toward hybrid approaches reflects the pragmatic response to current hardware limitations while still seeking to demonstrate quantum advantage where possible. As quantum hardware continues to improve, we can expect to see more ambitious QGAN implementations that leverage larger qubit counts and deeper circuits.

The realization of two-qubit controlled gates remains a major challenge in developing scalable quantum circuits with high fidelity. To address this, quantum Zeno dynamics has been employed for two-qubit entanglement generation~\cite{wang2018quantum} and for constructing quantum repeaters~\cite{bayrakci2022quantum} without relying on CNOT gates. Moreover, quantum Zeno dynamics has been used to reduce circuit complexity and to enable advantages in the activation~\cite{ozaydin2022nonlocal} and superactivation~\cite{ozaydin2023superactivating} of bound entanglement. Therefore, quantum Zeno dynamics-also referred to as quantum Zeno computation- may offer a promising approach for mitigating circuit complexity limitations in QGAN implementations.

\section{Recent Advances}
Since there have been extensive reviews on QGANs in early 2023 such as 
Ref.~\cite{ngo2023survey}, we focus on works since then.

In 2023, Silver et al. introduced MosaiQ~\cite{silver2023mosaiq} which a hybrid quantum-classical generative adversarial network (QGAN) framework designed for high-quality image generation on NISQ (Noisy Intermediate-Scale Quantum) devices. It introduces a scalable architecture composed of low-depth variational quantum circuits and utilizes a novel feature redistribution technique to effectively exploit principal components of image data-offering a significant improvement over pixel-by-pixel learning strategies. To address mode collapse and improve generative diversity, MosaiQ incorporates an adaptive input noise generation method. Experimental evaluations conducted on both simulators and real quantum hardware (e.g., IBM Jakarta) using standard benchmarks such as MNIST and Fashion-MNIST demonstrate that MosaiQ consistently outperforms previous state-of-the-art methods in both visual quality and quantitative metrics, including over 100-point improvements in Frechet Inception Distance (FID) scores. By making its implementation available as open-source, MosaiQ contributes a practical and robust QGAN framework to the emerging field of quantum image synthesis.

The potential of quantum machine learning has recently been explored in the context of cybersecurity, specifically for building more adaptive and efficient intrusion detection systems (IDS). One proposed approach leverages quantum generative adversarial networks (QGANs) to identify malicious network activity by learning atypical traffic patterns that are difficult to capture using classical models~\cite{rahman2023quantum}. By integrating QGANs within a framework such as IBM's Qiskit, the system can model and detect attack sequences at the protocol level using quantum-enhanced pattern recognition. Unlike classical intrusion detection systems that may suffer from high training costs or reduced accuracy on large datasets, the quantum approach offers an alternative path toward scalable anomaly detection with potential computational advantages. This application underscores the utility of QGANs in domains where dynamic threat landscapes require models capable of learning complex, high-dimensional behaviors.

A promising application of QGANs has emerged in the context of de novo drug design, where the integration of quantum computing into generative models aims to accelerate early-stage therapeutic discovery. Hybrid quantum-classical GAN architectures were employed to generate small drug-like molecules with enhanced physicochemical properties~\cite{kao2023exploring}. By replacing classical components of the GAN with variational quantum circuits (VQCs), the authors demonstrated improved performance on goal-directed benchmarks compared to traditional GANs. Notably, using a VQC in the generator allowed for the synthesis of higher-quality molecular structures using significantly fewer learnable parameters. The study further explored the use of quantum discriminators, where a minimal set of quantum parameters in the discriminator still led to better generative outcomes than multilayer perceptron (MLP)-based baselines, as measured by metrics such as KL divergence. Despite these advances, the hybrid models continue to face limitations in reliably generating diverse and chemically valid molecules, highlighting both the potential and current boundaries of QGAN-based approaches in generative chemistry.

A conditional adversarial approach was recently proposed to accelerate the modeling of ground-state properties in interacting quantum systems, particularly for predicting charge densities from confinement potentials in two-dimensional structures~\cite{pantis2023mapping}. Instead of relying on conventional variational wavefunction assumptions, the method uses a data-driven image-to-image translation framework to directly learn mappings between input potentials and resulting electron densities. This approach enables accurate forward predictions of both interacting and non-interacting systems, and intriguingly, also supports inverse design by reconstructing confinement potentials from target densities. The model performs well across various configurations and scales efficiently for a fixed class of systems, offering a viable alternative to computationally expensive exact diagonalization techniques. Beyond its predictive capabilities, the framework shows promise for solving inverse quantum problems and may facilitate device-level design tasks in mesoscopic and nanoelectronic systems.

A recent contribution proposed a fully quantum generative adversarial network architecture tailored for binary data, aiming to overcome limitations in earlier QGAN models that relied on hybrid configurations~\cite{chaudhary2023towards}. The design introduces several novel features, including a quantum generator that employs noise reuploading strategies and a quantum discriminator enhanced with auxiliary qubits for greater expressivity. A direct quantum interface between the generator and discriminator eliminates the need to estimate output probability distributions, streamlining the training process. The model leverages a Wasserstein-based loss function to improve training stability, addressing issues such as gradient vanishing often seen in adversarial learning. Empirical evaluations on synthetic datasets and low-energy configurations of the Ising model indicate that the architecture is capable of capturing and generalizing from discrete data distributions. This work moves closer to realizing scalable, fully quantum generative models by integrating advanced circuit design with principled training dynamics.

Adversarial vulnerability in quantum neural networks (QNNs) has emerged as a critical concern for the robustness of quantum machine learning systems. To address this, a recent study proposes enhancing the defense of QNN-based classifiers by embedding noise layers directly into the network architecture~\cite{huang2023enhancing}. Unlike traditional methods that add noise only to the input data during training, this approach applies noise consistently throughout both training and inference phases, allowing the model to learn more resilient parameters and resist gradient-based attacks more effectively. The noise strength is governed by a min-max optimization scheme, enabling adaptive regularization without significantly compromising accuracy on clean data. Comparative experiments demonstrate that this method outperforms existing defense techniques, including quantum adversarial training and differential privacy-based approaches. While the approach incurs greater computational costs, it offers a promising direction for building more robust and generalizable quantum classifiers, advancing the reliability of quantum AI applications under adversarial conditions.

To support the development of AI-driven diagnostic systems in cardiology, a conditional quantum generative framework has been proposed for synthesizing abnormal electrocardiogram (ECG) signals~\cite{qu2023quantum}. Addressing the challenges of class imbalance and limited data availability in abnormal ECG datasets, the model adopts a patch-based generation strategy, where multiple sub-generators are employed to construct different heartbeat features across signal segments. This modular design conserves quantum resources and enhances compatibility with near-term devices. Furthermore, the model incorporates quantum registers that encode condition information, such as heartbeat types and their distributions, enabling controlled generation of specific abnormalities. Simulations reveal that the framework achieves high accuracy in reproducing abnormal patterns and demonstrates notable robustness under quantum noise, suggesting practical value in medical data augmentation tasks. This work highlights the applicability of QGAN architectures in healthcare, particularly in contexts where data diversity and controllability are essential.

In 2024, Hamiltonian Quantum Generative Adversarial Networks (HQuGANs) introduce a novel framework for learning unknown quantum states by leveraging two competing quantum optimal control strategies instead of the traditional circuit-based architecture~\cite{kim2024hamiltonian}. This formulation allows the model to operate under continuous-time Hamiltonian dynamics, making it naturally compatible with experimental constraints such as bandwidth-limited control fields. The approach is particularly well-suited for generating highly entangled many-body states and offers an expanded unitary search space compared to gate-based methods. Numerical simulations confirm the effectiveness of HQuGANs using GRAPE and Krotov's optimization techniques, and the proposed cost function-based on the quantum Wasserstein distance-helps mitigate mode collapse and accelerates convergence. The framework also extends beyond pure state generation to learning quantum processes, positioning it as a flexible tool for near-term quantum platforms including superconducting and ion-trap systems. This work highlights how integrating game-theoretic principles with quantum control can enable scalable, noise-resilient generative learning in quantum settings.

ReCon was introduced as a novel QGAN implementation on analog Rydberg atom quantum computers, marking a departure from earlier efforts primarily based on superconducting-qubit platforms~\cite{dibrita2024recon}. Capitalizing on the reconfigurable qubit layout, multi-qubit interaction capability, and longer coherence times of Rydberg atom systems, ReCon demonstrates a hybrid architecture with a quantum generator and classical discriminator. The generator is optimized through dimensionality reduction and customized pulse-based control over qubit evolution, enabling image generation using just four qubits. Through layered training and ensemble learning techniques, ReCon achieves improved sample quality and diversity. Empirical evaluations using benchmark image datasets show a 33\% improvement in Frechet Inception Distance (FID) over prior state-of-the-art superconducting-based QGANs, in particular MosaiQ~\cite{silver2023mosaiq}. Notably, ReCon is also validated on real hardware via AWS Braket using the QuEra Aquila device, showcasing its practical feasibility. This work illustrates the potential of alternative quantum hardware platforms in pushing QGAN capabilities forward, particularly under noisy and resource-constrained conditions.

To address the persistent challenge of readout errors in quantum circuits, a QGAN framework incorporating a bit-flip averaging (BFA) strategy has been proposed for enhanced fault tolerance during image generation tasks~\cite{zhao2024quantum}. This method simplifies the error mitigation process by pre-averaging qubit measurements across random bit-flip events, thereby avoiding the computational overhead of traditional response matrix techniques. Simulations conducted on Qiskit using a handwritten digit dataset demonstrate that the model maintains high fidelity and low divergence across varying readout error rates, with KL divergence values remaining below 0.1 and state fidelities exceeding 0.95 even at higher noise levels. The results confirm the effectiveness of BFA in stabilizing QGAN performance under noisy conditions, offering a practical error mitigation pathway for image-based quantum generative models on NISQ hardware.

The implementation of quantum generative adversarial networks (QGANs) was demonstrated on a silicon photonic quantum chip~\cite{ma2024quantum}. The two-qubit photonic device enables arbitrary controlled-unitary operations and can generate any two-qubit pure state, making it well-suited for generative learning. To enhance model capacity, a hybrid generator introduces classical nonlinearity into the QGAN architecture. Three benchmark tasks were successfully executed on the chip: single-qubit pure state learning, classical probability distribution loading, and compressed image generation. These results validate the potential of photonic quantum hardware in executing both pure and hybrid QGANs with practical generative capabilities, reinforcing its promise for future scalable quantum machine learning applications.

A photonic QGAN was designed to generate classical image data using linear optical circuits and Fock-space encoding~\cite{sedrakyan2024photonic}. Unlike qubit-based QGANs, the model is implemented on a photonic processor (Quandela's Ascella) and trained end-to-end to produce digit-like images. The quantum generator employs variational circuits, and experimental results show convergence of the loss function and visually coherent outputs, despite limited diversity.
The architecture combines noise reuploading and patch-based learning to support higher-resolution outputs. The use of Fock space provides a resource-efficient alternative to qubit encoding, with fast sampling and favorable scaling on photonic hardware. Optimization via SPSA proved effective even under noisy conditions. Future directions include extending the model to conditional QGANs and applying it to domains like high-energy physics and quantum chemistry, where structured classical data may benefit from quantum generative modeling.

An evolutionary quantum generative adversarial network (EQGAN) was introduced as a novel framework that enhances standard QGAN performance by incorporating evolutionary strategies~\cite{xie2024evolutionary}. Unlike conventional QGANs that often suffer from limited adversarial objectives and risk of mode collapse, EQGAN leverages multiple adversarial training targets as mutation operations. Quantum circuits serve as generators while classical neural networks act as discriminators. A fitness evaluation is employed to select optimal generator configurations across generations, improving convergence speed and reducing training costs. Simulations conducted on two datasets demonstrate that EQGAN achieves faster and more stable training compared to traditional approaches, effectively addressing limitations related to adversarial singularity.

Integration of quantum computational elements into classical Generative Adversarial Networks was investigated for advancing the frontier of quantum-enhanced machine learning~\cite{nokhwal2024quantum}. By incorporating quantum data representation techniques and leveraging the distinctive capabilities of qubits, the study proposes a hybrid GAN architecture intended to accelerate convergence and improve generative quality. Theoretical and experimental considerations are discussed, including the potential quantum advantages in training speed and data representation fidelity. The work also addresses practical challenges such as hardware limitations, error correction, and scalability. Positioned at the intersection of classical and quantum machine learning, the study contributes to ongoing efforts to harness quantum computational power for improving the performance of generative models.

To address the common issue of mode collapse in QGANs, an unrolled QGAN model within a hybrid quantum-classical framework was proposed~\cite{gong2024unrolled}. The approach involves unrolling the training process by separately optimizing the discriminator over multiple steps before updating the generator. This unrolling strategy helps stabilize training and leads to more robust generator updates. The model is evaluated on tasks involving quantum and Gaussian distribution generation, with performance metrics including KL divergence, standard deviation, and mean value. Both numerical simulations and experimental results demonstrate that the unrolled QGAN enhances data diversity and improves coverage of the target distributions, achieving more reliable generative outcomes than conventional QGANs.

A series of hybrid QGAN models have been developed to improve the efficiency and accuracy of small organic molecule generation in computational drug discovery~\cite{cui2024efficient}. Building on earlier work that applied QGANs to generate drug-like compounds, the authors introduced QNetGAN, an adaptation inspired by classical NetGAN, which generates molecular graph structures via random walks. This approach demonstrated substantial improvements in both speed and success rate over the original QGAN model. The latest version, QNetGAN v2, incorporates graph convolution techniques and multiple post-processing steps, including rule-based filters for chemical validity and geometric consistency-resulting in a 91\% success rate in generating valid molecular structures. With a notably reduced training time of just 35 minutes, QNetGAN v2 marks a significant step toward practical quantum-assisted molecular design. These developments illustrate the evolving role of QGANs in graph-based molecular generation and their growing promise for accelerating early-stage drug discovery under resource constraints.
HQ-MolGAN is another hybrid framework tailored for small molecule generation in drug discovery~\cite{anoshin2024hybrid}. By integrating parameterized quantum circuits into the generator architecture and introducing a cycle-consistency mechanism, the model enhances stability and performance during training. Experimental evaluations on benchmark datasets demonstrate substantial improvements over classical and earlier hybrid models, including up to a 30\% increase in drug-likeness metrics. The model also shows strong resilience to quantum noise, performing comparably well in both ideal and noisy simulation environments. Notably, HQ-MolGAN achieves competitive results using fewer training iterations and reduced model complexity, highlighting its efficiency. The inclusion of a hybrid cycle component further mitigates issues like high-entropy states and improves molecular uniqueness. This work underscores the potential of hybrid quantum generative models in pharmacological applications and sets the stage for future improvements in model architecture and dataset diversity aimed at accelerating the drug development process.

Another hybrid quantum-classical GAN framework was designed for near-term quantum devices, emphasizing scalability and circuit efficiency~\cite{boyle2024hybrid}. The generator is implemented via variational quantum circuits with angle encoding, while the discriminator uses a modular, multi-stage quantum neural network design that enables fine-grained control over circuit depth and complexity. A key innovation in this work lies in computing gradients directly from the same quantum circuits used for network operations, avoiding the overhead of auxiliary circuits or qubits. Simulated using IBM's Qiskit and trained with classical mini-batch stochastic gradient descent, the model demonstrates effective performance on two-qubit systems. Notably, a five-stage discriminator configuration yields favorable divergence scores, showing promising alignment between generated and real data distributions. This modular, resource-aware design highlights a viable path for implementing QGANs on current quantum hardware with constrained capabilities, balancing practical training demands with performance objectives.

A novel enhancement to classical super-resolution GANs has been proposed by incorporating a quantum-inspired feature layer, motivated by the principle of quantum superposition~\cite{pu2024super}. This augmented architecture is applied to aerial agricultural images, aiming to reconstruct high-resolution images from low-resolution inputs with improved fidelity. By modifying the baseline super-resolution GAN (SRGAN) model, the approach integrates a custom quantum feature enhancement layer derived from the pix2pix architecture, which is a conditional GAN designed for image-to-image translation tasks~\cite{qin2023laser}. Empirical evaluations on UAV-collected image datasets demonstrate consistent improvements in structural similarity (SSIM) and peak signal-to-noise ratio (PSNR), achieving an 8\% gain in reconstruction quality over the original SRGAN. The model's effectiveness is further supported by t-SNE visualizations and experiments on high-resolution public datasets. This work illustrates how hybrid classical-quantum design elements can improve generative image processing tasks, particularly in high-precision applications like agricultural monitoring.

The use of quantum-enhanced generative models was explored for financial time series synthesis, specifically targeting the statistical replication of the S\&P 500 index~\cite{orlandi2024enhancing}. The proposed approach, termed QWGAN-GP, combines a quantum generator with a classical discriminator and integrates a gradient penalty to stabilize training. Evaluations using metrics such as Wasserstein distance, dynamic time warping, and entropy measures confirm the high fidelity of the generated sequences. Beyond generation, the study assesses practical utility by incorporating synthetic data into long term short memory based forecasting models. Results show that blending quantum-generated and real data leads to notably improved predictive accuracy, especially in capturing market trends and rare fluctuations. This case demonstrates the potential of QGAN-based models to enhance financial modeling and risk assessment by providing high-quality synthetic training data.

A novel quantum computing framework has been proposed for estimating financial risk in carbon markets, combining a quantum conditional generative adversarial network (QCGAN) with quantum amplitude estimation (QAE)~\cite{zhou2024carbon}. The hybrid QCGAN-QAE model generates return rate distributions and efficiently computes risk metrics such as value-at-risk (VaR) and conditional value-at-risk (CVaR), without relying on strong distributional assumptions. Key innovations include a redesigned quantum generator circuit that reorders data entanglement and simulation layers, and the introduction of a quantum fully connected layer to enhance model expressivity. Additionally, a binary search strategy integrated into QAE boosts computational efficiency by narrowing amplitude search space. Simulation results based on the EU Emissions Trading System show that this framework significantly improves both accuracy and performance over classical methods, positioning it as a promising approach for quantum-enhanced financial risk analysis.

To address class imbalance in ECG datasets-a common barrier in developing robust diagnostic tools-HQ-DCGAN introduces a hybrid quantum-classical GAN architecture tailored for synthetic ECG signal generation~\cite{qu2024hq}. The model integrates quantum convolutional layers and parameterized quantum circuits into both generator and discriminator networks, enhancing feature extraction and leveraging quantum resources efficiently. It maintains the expressive power and scalability of classical deep convolutional models while remaining compatible with NISQ-era constraints on qubit count and circuit depth. A custom evaluation metric, the 1D Frechet Inception Distance (1DFID), is proposed to assess signal quality. Simulation results demonstrate that HQ-DCGAN produces high-fidelity ECG waveforms and achieves superior classification accuracy (82.2\%) compared to baseline methods. This work highlights the viability of hybrid quantum convolutional architectures in biomedical signal synthesis under current hardware limitations.

In 2025, an optimized self-guided quantum generative adversarial network (SGQGAN) framework was proposed~\cite{selvam2025optimized}, which is enhanced by the Prairie Dog Optimization Algorithm (PDOA), for efficient task scheduling and resource allocation in cloud computing environments. Aiming to overcome challenges in virtual machine (VM) interference and performance degradation, the SGQGAN-PDOA framework integrates Inception Transformer (IT) for robust feature extraction and Spatial Distribution-Principal Component Analysis (SD-PCA) for dimensionality reduction. Implemented on Java and CloudSim, the system dynamically reallocates tasks to improve utilization, achieving 80\% reliability with 150 VMs at 200 ms processing latency. The hybrid quantum-classical approach significantly reduces waiting time, response time, and load imbalance compared to existing scheduling methods. Results highlight the scalability and responsiveness of quantum-inspired resource management, positioning SGQGAN-PDOA as a promising model for enhancing performance and reliability in dynamic cloud infrastructures.

By utilizing Large Language Models (LLMs), a novel method for optimizing ansatz design in QGANs was proposed~\cite{ueda2025optimizing}. The approach integrates LLMs into an iterative refinement loop to enhance ansatz structures, aiming to improve generative accuracy while minimizing circuit depth and parameter count. Beyond qGANs, the framework is applicable to a broad class of variational quantum algorithms, offering a flexible and generalizable strategy for AI-assisted quantum circuit optimization. This work highlights the emerging role of LLMs in automating and advancing quantum algorithm development.

An end-to-end implementation of a hybrid quantum-classical GAN for augmenting electron backscatter diffraction (EBSD) images of steel microstructures was demonstrated~\cite{sekwao2025end}. By integrating a quantum circuit into the input layer of a classical Wasserstein GAN (WGAN), the model generates synthetic images of ferrite and bainite phases, addressing data scarcity in materials science. The approach demonstrates improved image quality and reduced mode collapse, outperforming classical Bernoulli GANs in 70\% of cases based on maximum mean discrepancy scores. The results suggest the potential of quantum-assisted generative models for enhancing image datasets in scientific applications, with scalability prospects as quantum hardware advances.

HyperKING, a novel hybrid quantum-classical generative adversarial network (GAN) architecture was designed for large-scale hyperspectral image (HSI) restoration in satellite remote sensing (SRS)~\cite{lin2025hyperking}. Addressing the limitations of prior quantum GANs constrained by qubit resources, often limited to generating $2\cross2$ or $28\cross28$ grayscale images, HyperKING innovatively integrates hybrid generator and discriminator networks. The quantum components are engineered for full expressibility, allowing the realization of any valid quantum operator via training, while classical convolutional layers manage the encoding and decoding processes to interface with the quantum domain. The model effectively mitigates quantum collapse effects and supports processing of $128\cross 128$ HSIs. Experimental results show that HyperKING outperforms classical methods in hyperspectral tensor completion, mixed noise removal (achieving $\approx3$ dB PSNR gain), and blind source separation, marking a significant advancement in quantum-enhanced image restoration for real-world SRS applications.

Accurate channel prediction is essential for optimal performance in large-scale multiple-input multiple-output orthogonal frequency division multiplexing (MIMO-OFDM) systems. A hybrid quantum deep convolutional generative adversarial network (HQDCGAN) framework, referred to as HQDCGAN-MIMO-OFDM, was introduced for enhancing channel prediction accuracy and addressing performance bottlenecks such as error vector magnitude (EVM), peak-to-average power ratio (PAPR), and adjacent channel leakage ratio (ACLR)~\cite{vijayakumari2025hybrid}. The HQDCGAN leverages pyramidal dilated convolutions and attention mechanisms to extract multi-scale features from OFDM channel data, enabling the model to capture complex spatial-temporal correlations. A PAPR reduction module, trained on lower PAPR data via a simplified clipping with filtering (SCF) technique, further enhances efficiency. The model is evaluated using key metrics such as spectral efficiency, PAPR, bit error rate (BER), signal-to-noise ratio (SNR), and throughput. Comparative analysis demonstrates that the proposed HQDCGAN-MIMO-OFDM method outperforms conventional methods significantly, confirming its potential for robust and high-performance wireless communication systems.

A novel educational tool that combines QGANs and transformer-based models was proposed to enhance learning through interactive storytelling~\cite{sriraksha2025quantales}. Aimed at addressing the limitations of traditional Indian school systems-which often rely on rote memorization-this application enables students and teachers to generate visual, narrative-based learning content from curriculum-aligned prompts. The model extracts key concepts from user inputs and builds dynamic story graphs to support conceptual understanding. By integrating quantum computing, the system boosts generative capabilities, offering a creative and engaging educational experience that supports both self-study and classroom instruction.

A quantum image generative learning (QIGL) approach was proposed to overcome the limitations of conventional QGANs in producing high-resolution medical images~\cite{khatun2025quantum}. Unlike patch-based pixel-wise learning models that fail to capture global structures, QIGL employs a variational quantum circuit that extracts principal components from full images, improving scalability and feature representation. The model also incorporates the Wasserstein distance to enhance the diversity and quality of generated medical samples. Experiments on knee osteoarthritis and medical MNIST x-ray datasets show that QIGL outperforms both classical GANs and existing QGANs, achieving notably lower Frechet Inception Distance scores.

A quantum-inspired model called DBO-QH-GAN was designed for enhanced software defect prediction that addresses key limitations in current approaches, such as feature redundancy, noise, and dataset bias~\cite{chaudhary2025enhanced}. The method begins by extracting software metrics from PROMISE datasets, followed by a Fuzzy K-Top Matcher (FK-TM) step to clean the data by removing noise, outliers, and redundant features. The Botox Optimization algorithm then selects five critical software metrics, including cyclomatic complexity, Halstead measures, and code churn. These refined features are input into a Quantum Hamiltonian Generative Adversarial Network~\cite{kim2024hamiltonian}, which identifies software defects using probabilistic thresholding. The process is further refined using Duck Bevy Optimization (DBO), enhancing model performance, offering a highly reliable and cost-effective solution for software defect prediction.

\section*{Conclusion}
Quantum Generative Adversarial Networks represent a convergence of two transformative technologies: quantum computing and generative modeling. This review examined the diverse range of QGAN architectures, from hybrid quantum-classical frameworks to fully quantum circuits implemented on near-term hardware. Across domains including medical imaging, materials science, software reliability, and education, QGANs have shown potential for enhancing data generation tasks, improving performance, and introducing novel design strategies such as quantum ansatz optimization via language models.

Despite promising early results, several challenges remain before QGANs can achieve widespread practical deployment. These include limitations in quantum hardware scalability, optimization convergence, and training stability, as well as the need for standardized benchmarks across quantum platforms. Nonetheless, the flexibility of QGAN architectures, their capacity to encode complex data distributions, and ongoing improvements in quantum processors point toward a future where QGANs could play a central role in quantum machine learning.

Future work will likely focus on conditional QGANs, improved variational circuit design, robust error mitigation strategies, the generation of genuinely quantum data, as well as algorithmic quantum fairness~\cite{chintalapati2024quantum}. As quantum technologies continue to evolve, QGANs stand as a compelling candidate for demonstrating early quantum advantage in real-world applications.

\bmhead{Acknowledgements}

F.O. acknowledges Tokyo International University Personal Research Fund.

\section*{Declarations}

\begin{itemize}
	\item Funding: Tokyo International University Personal Research Fund.
	\item Conflict of interest/Competing interests: Not applicable.
	\item Ethics approval and consent to participate: Not applicable.
	\item Consent for publication: Not applicable.
	\item Data availability: No data is generated in this work.
	\item Materials availability: Not applicable.
	\item Code availability: Not applicable. 
	\item Author contribution: M.I., S.T. and F.O. conceptualization, research, manuscript preparation. S.T. and F.O., supervision. All authors have viewed and agreed on the final manuscript.
\end{itemize}

\bibliography{OQuLv20250522}
\end{document}